\documentclass[12pt,preprint]{aastex}
\usepackage{url}
\shorttitle{Evershed Effect in Sunspots}
\title{Traveling Waves of Magnetoconvection and the Origin of the Evershed Effect in Sunspots}

\author{I. N. Kitiashvili}
\affil{Center for Turbulence Research, Stanford University, Stanford, CA 94305, USA}
\email{irinasun@stanford.edu}
\author{A. G. Kosovichev}
\affil{Hansen Experimental Physics Laboratory, Stanford
University, Stanford, CA 94305, USA}
\email{sasha@sun.stanford.edu}
\author{A. A. Wray}
\affil{NASA Ames Research Center, Moffett Field,Mountain View, CA 94040, USA}
\email{A.A.Wray@nasa.gov}
\and
\author{N. N. Mansour}
\affil{NASA Ames Research Center, Moffett Field,Mountain View, CA 94040, USA}
\email{N.N.Mansour@nasa.gov}
\begin{document}

\begin{abstract}
Discovered in 1909 the Evershed effect represents strong mass
outflows in sunspot penumbra, where the magnetic field of sunspots
is filamentary and almost horizontal. These flows play important
role in sunspots and have been studied in detail using large
ground-based and space telescopes, but the basic understanding of
its mechanism is still missing. We present results of realistic
numerical simulations of the Sun's subsurface dynamics, and argue
that the key mechanism of this effect is in non-linear magnetoconvection
that has properties of traveling  waves in the presence of strong, highly inclined
magnetic field. The simulations reproduce many observed features of
the Evershed effect, including the high-speed ''Evershed clouds",
the filamentary structure of the flows, and the non-stationary
quasi-periodic behavior. The results provide a synergy of previous
theoretical models and lead to an interesting prediction of a
large-scale organization of the outflows.

\end{abstract}

\section{Introduction}
In the spring of 1909 Evershed published a
remarkable discovery of strong horizontal mass flows in sunspots
penumbra, the outer part of sunspots characterized by filamentary
magnetic field structures  \citep{Evershed1909}. The flows, with typical speed of 1--4
km/s, start at the boundary between the umbra and penumbra and
expand radially, accelerating with distance and suddenly stopping at
the outer sunspot boundary.  The Evershed effect may play a
significant role in the formation, stability and dynamics of
sunspots and is considered as one of the fundamental process in
solar physics. This phenomenon caused significant interest and
detailed observational and theoretical investigations, but the
understanding of the physical mechanism is still missing (a
recent review is published by  \citet{Tritschler2009}).

High-resolution observations from large ground-based telescopes and
the Hinode space mission revealed a complicated filamentary
structure of these
flows \citep{Rimmele1994,Rimmele1995,Ichimoto2007a,Ichimoto2007b} and
their non-stationary dynamics in the form of quasiperiodic
''Evershed
clouds" \citep{Shine1994,Rimmele1994,Georgakilas2003,CabreraSolana2007,CabreraSolana2008}.
In some cases, the flows showed a large-scale coherent behavior
across several flow channels \citep{Shine1994}, and also provided
evidence of a wave-like behavior \citep{Rimmele1994,Georgakilas2003}.

Theories of the Evershed effect can be divided in two categories,
describing it as channel flows in magnetic flux
tubes \citep{Meyer1968,Montesinos1997,Schlichenmaier1998} or as
elongated magnetoconvective rolls \citep{Danielson1961,Hurlburt2000}.
Recent numerical simulations \citep{Heinemann2007,Rempel2009}
successfully modeled the filamentary magnetic structure of sunspot
penumbra and horizontal outflows, thus providing a strong support to
the magnetoconvective nature of the Evershed
effect \citep{Scharmer2008}.

In this paper, we present a study of solar magnetoconvection in the
presence of inclined magnetic field, based on the realistic
radiative MHD simulations, and link the Evershed effect to the
phenomenon of traveling magnetoconvection waves. The convective
waves is a very interesting MHD phenomenon \citep{Weiss1991}, which, in
fact, has been previously suggested as a reason of the Evershed
flows \citep{Hurlburt2000}, but did not receive further development.
Our study provides a basis for explaining the Evershed effect as a
result of traveling magnetoconvection waves in highly inclined
magnetic field of sunspot penumbra. The phenomenon of traveling
magnetoconvection waves is considered also in the dynamics of the
Earth's core \citep{Walker1999,Zhang1999}, and may happen in various
astrophysical objects, such as magnetic stars, accretion disks,
compact objects and active galactic nuclei. Thus, detailed
observational and theoretical studies of this phenomenon are of
great interest.

\section{Numerical Simulations}

We use a three-dimensional non-linear radiative-magnetohydrodynamics
code developed for simulating the upper solar convection zone and
lower atmosphere  \citep{Jacoutot2008a,Jacoutot2008b}. This code takes
into account several physical phenomena: compressible fluid flow in
a highly stratified medium, 3D multi-group radiative energy transfer
between the fluid elements, a real-gas equation of state, ionization
and excitation of all abundant species, and magnetic effects. A
important feature of this code is implementation of various subgrid
scale turbulence models. In this paper, we adopted the most widely
used Smagorinsky model  \citep{Smagorinsky1963} in the
compressible formulation  \citep{Moin1991,Germano1991}. The turbulent
electrical conductivity is calculated by using the extension of the
Smagorinsky model to the MHD case  \citep{Theobald1994}.

We simulate the upper layer of the convection zone, extending from
5~Mm below the visible surface to 0.5 Mm above the surface. The
horizontal size varied from 6.4~Mm~$\times$~6.4~Mm to 25~Mm~$\times$
25~Mm. The computational grid step size varied from 25 to 100 km. The results
of this paper are obtained using $128^3$ grid with the step size of 50 km
(except Fig.~5, which is obtained using $64^3$ grid). We have verified that the
results do not change in computations with finer grids (except of the development
of smaller scale turbulent motions).

The initial uniform magnetic field is imposed on a snapshot of the
preexisting hydrodynamic convection  \citep{Jacoutot2008b}. The
initial field strength, $B_0$, varies from 0 to 2000 Gauss, and the
inclination angle, $\alpha$, varies from 0 to 90 degrees. The lateral boundary
conditions are periodic, and the top and bottom boundary conditions
maintain the total magnetic flux and the mean inclination. This
formulation allows us to carry out a series of controlled numerical
experiments and investigate how the structure and dynamics of solar
turbulent convection depend on the magnetic field properties in
regimes close to the observed in sunspot penumbra, and elucidate the
physical mechanism of the Evershed effect.

\section{Results}

Outside magnetic field regions the solar convection forms granular
cells of a typical size of 1-2 Mm and lifetime of about 10 min. In
the presence of magnetic field the structure of convection strongly
depend on the field strength and inclination. When the magnetic
field is vertical the granules become smaller \citep{Stein2002}, and their overturn
time is shorter resulting in generation of high-frequency turbulence
and acoustic waves ("halos")  \citep{Jacoutot2008b}. In the presence
of an inclined magnetic field, such as observed in sunspot penumbra,
the granular cells becomes naturally elongated in the direction of
the field because magnetic field restricts motions across the
magnetic field lines. But the most interesting effect is that the
inclined field changes the nature of solar convection. Instead of a
stationary overturning convection pattern the simulation reveal
traveling convection waves, which become more apparent and stronger
for higher field strengths and inclination. This convection develops
long narrow structures of velocity, thermodynamic parameters and
magnetic field, resembling the filamentary structure and motions in
the penumbra of sunspots. These structures are illustrated in
Figures 1 and 2.

Figure 1 shows a 3D slice of our computational domain with a sample of
magnetic field lines (red curves), velocity field (black arrows) and
a volume rendering of the temperature structures (blue-red color scale).
The initial 1000 Gauss magnetic field is oriented in the $xz$-plane
and inclined by 85 degrees to the $z$-axis, so that the $B$-vector
is the positive in the $x$-direction. Evidently the strongest
motions occur in the direction of the field inclination in narrow
structures, with upflows and downflows at the initial and end point
of these structures.  The magnetic field lines
change in accord with these elongated motions, rising up at the
initial points and declining at the end points, thus giving an
impression of rising and falling loop-like motions. The temperature is typically higher
at the start points and lower at the end points. The typical
vertical velocity around these points is about 1 km/s, but the
horizontal velocity between them in the positive $x$-direction
reaches 4--6 km/s. Most of the horizontal mass flow occur in these
relatively narrow patches, which strongly resemble ''Evershed
clouds", discovered in observations \citep{Shine1994}. Significantly
weaker flows in the opposite direction are also observed. These are often
originate at the initial upflow points.
Vertical cuts through the flow field (e.g. left $yz$-plane
in Fig.~1) reveal associated vortex-type motions below the surface.

When the background field is strong, the horizontal flow patches
become quite narrow, with the width of 0.5 Mm or less. The magnetic
field variations also becomes more filamentary. This is illustrated
in Fig.~2, which shows the surface structure of the horizontal flows
and the $B_x$ component for two different initial magnetic field strengths,
1000 and 1200 G. The simulations show strong interaction between the plasma flows and
magnetic field. The magnetic field controls the general direction of
the flows, but in the strong flow patches, the magnetic field is
pushed aside and has a reduced magnetic field strength. This may
give impression, sometimes, reported from observations that the
flows occur in magnetic field ''gaps". Nevertheless, the plasma
flows remain magnetized. The filamentary magnetic structures and
 flows are strongly coupled.

The most interesting feature of the simulations, which, we argue, is
a key for understanding the Evershed effect, is the traveling wave
pattern of magnetoconvection in the presence of strongly inclined
magnetic field. The simulations show that the velocity patches and
magnetic field perturbations migrate in the direction of the field
inclination. Vertical cuts in $xz$-planes  show
rapidly moving inclined convective cells (a snapshot is illustrated in Fig.~3).
This process is best seen in the movies\footnote{\url{
http://soi.stanford.edu/~irina/Simulations/movies.html}}, and also in the
time-distance slices of the surface $V_x$ velocity component
along the $x$-axis. Figure 4 shows an example of these slices for the
initial magnetic field, $B_0=1200$ G, and the inclination angle
of $85^\circ$. In this case the convective velocity reaches $\sim 6$ km/s,
and a pattern of convection waves traveling in the direction of
the field inclination with a speed of 1-2 km/s can be identified.

The general picture is that the overturning convection
motions are swept by the traveling waves. This interaction amplifies
the flows in the direction of the waves. This is accompanied by
weaker plasma motions in the opposite direction. In fact, the
initial points of the convective upflows often move in the opposite
direction. This may explain the puzzling discrepancy between the
outward flow direction and the apparent motion of ''penumbra
grains".

It is intriguing that the traveling convection pattern shows
variations  with a characteristic time of 20-50 min, resembling the
quasi-periodic behavior noticed in the observations
 \citep{Shine1994,Rimmele1994,Georgakilas2003}. By increasing the
computational domain up to 25 Mm we have checked that the
quasi-periodicity does not depend on the size of the domain and,
thus, is not due to the periodic boundary condition. This is an
intrinsic property of the inclined field magnetoconvection, but
understanding of this phenomenon require further investigation.

As we have pointed out the high-speed (4-6 km/s) horizontal flows
occur in localized patches corresponding  to the ''Evershed clouds".
An averaged over time and space velocity is smaller, about 1--2
km/s. The flows are concentrated in a shallow subsurface layer less
than 1 Mm deep (Fig.~5a). The velocity peaks about 100--200 km below
the surface. This also corresponds to the observations showing that
the velocity of the Evershed flows increases with depth. The averaged
velocity does not change much with the magnetic field strength in the range
of 1000--1500 G, but it strongly depends on the inclination angle (Fig.~5b).
The mean horizontal flow is much weaker for small inclination angles.

\section{Discussion}

The radiative MHD simulations of solar magnetoconvection in regions
of inclined magnetic field qualitatively and quantitatively describe
many observed features of the Evershed effect in sunspots. The
results indicate that the principal physical mechanism of the
Evershed flows is the traveling wave nature of the
magnetoconvection. The traveling waves have been extensively studied
in idealized situations  \citep{Weiss1991,Hurlburt1996}, and it has
been suggested that they play significant role in sunspot flows
 \citep{Hurlburt2000}.  Our simulations model this phenomenon in the
realistic solar conditions, and show that indeed many details
correspond to the observations, thus providing a basis for
explaining the Evershed effect.

In particular, the simulations show that the high-speed flows
reaching 4-6 km/s occur in the direction of the field inclination in
narrow, 2--3 Mm long patches, which tend to appear quasi-periodically
on the time-scale of 15-40 min. These patches correspond to the
so-called ''Evershed clouds"
 \citep{Shine1994,Rimmele1994,CabreraSolana2007} and represent the
main component of the Evershed flows. These horizontal flows
originate from convective upflows of hotter plasma, like in ordinary
convection, but are channeled by the magnetic field and amplified by
the traveling convective waves. The whole process is highly
non-linear and stochastic with high-speed patches appearing
randomly, but the simulations also show large-scale organization
patterns across the simulation domain, which seem to be associated
with the traveling waves. These patterns are evident in the
simulation movies. Some observations showed a signature
of coherence in appearance of the Evershed clouds  \citep{Shine1994}
but this has not been fully
established \citep{Georgakilas2003}. The simulations suggest that a
large-scale coherence may be a fundamental property of the traveling
wave phenomenon, and certainly encourage further observational
studies. Of course, in real sunspots the magnetic field stricture is
highly inhomogeneous, and this may affect the large-scale
appearance. This must be investigated in future simulations.

In the past several models were suggested to explain the Evershed
effect. Interestingly, some features of these models can be found in our
simulations. One of the first models describes the penumbra
filaments as convective rolls along the direction of magnetic field
 \citep{Danielson1961}, first suggesting the convective nature of the
Evershed effect. The apparent observed wave-like behavior inspired
attempts to explain the Evershed effect as magnetoacoustic or
magnetogravity waves  \citep{Maltby1967,Bunte1993}. Our model
specifies that these waves are convective in nature. The rising and
falling thin-flux tube model \citep{Schlichenmaier1998} was suggested
to describe the discrepancy between the apparent motion of penumbra
features and the main Evershed flows. Our simulations explain this
naturally, and also reveal upward and downward loop-like motions of
magnetic field lines synchronized with the high-velocity patches.
The siphon model  \citep{Meyer1968,Montesinos1997} suggested that the
flow is driven by the pressure difference between the initial and
end points, and indeed, in the simulations the gas pressure in the
initial points is higher than at the end points. The recent
numerical simulations of the sunspot structure
 \citep{Heinemann2007,Rempel2009} led to the suggestion that the
Evershed effect is caused by the overturning convection
 \citep{Scharmer2008}, but the flow speed was not sufficiently high.
Our simulation show that the high-speed matching the observations is
achieved if the magnetic field is strong, 1000-1500 G, and highly
inclined, when the magnetoconvection has properties of traveling
waves. Thus, it seems that the MHD simulations provide a unified
description of the models and the key observed features, and,
perhaps, lead to the understanding of the 100-year old discovery.

\clearpage

\begin{figure}
\begin{center}
\includegraphics[width=\linewidth]{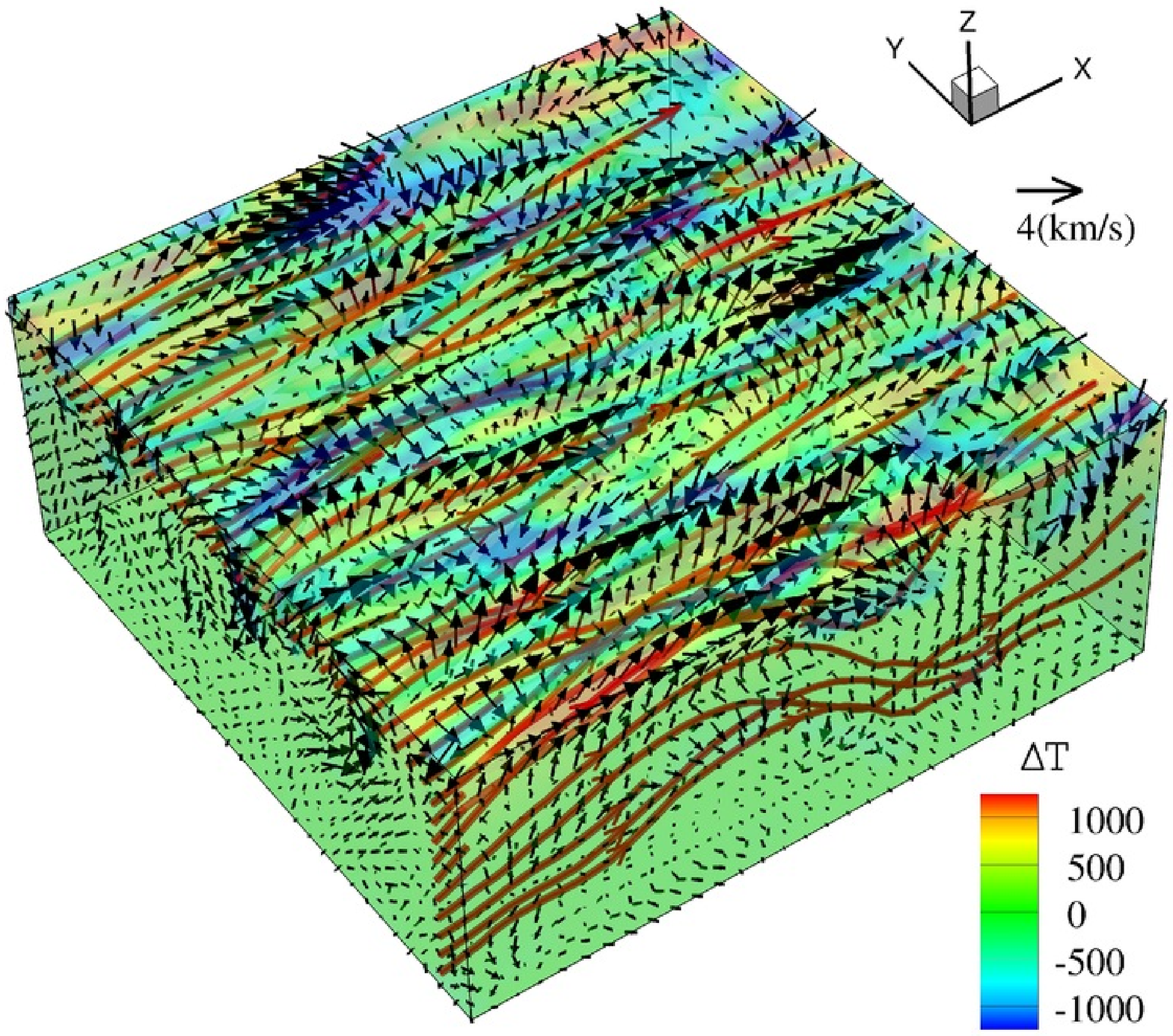}
\caption{A snapshot of the 3D simulations of subsurface solar
convection in a region of an inclined magnetic field. The initial
field strength, $B_0$, is 1000 G, the inclination angle, $\alpha$,
is 85 degrees with respect to the vertical. Red curves show magnetic
field lines; the color 3D structures show the temperature variations
relative to the mean vertical stratification
(dark green and blue correspond to lower temperature); the arrows
show the velocity field. The horizontal size of the box is 6.4 Mm,
the depth is about 2.5 Mm.}
\end{center}
\end{figure}

\begin{figure}
\begin{center}
\includegraphics[width=0.85\linewidth]{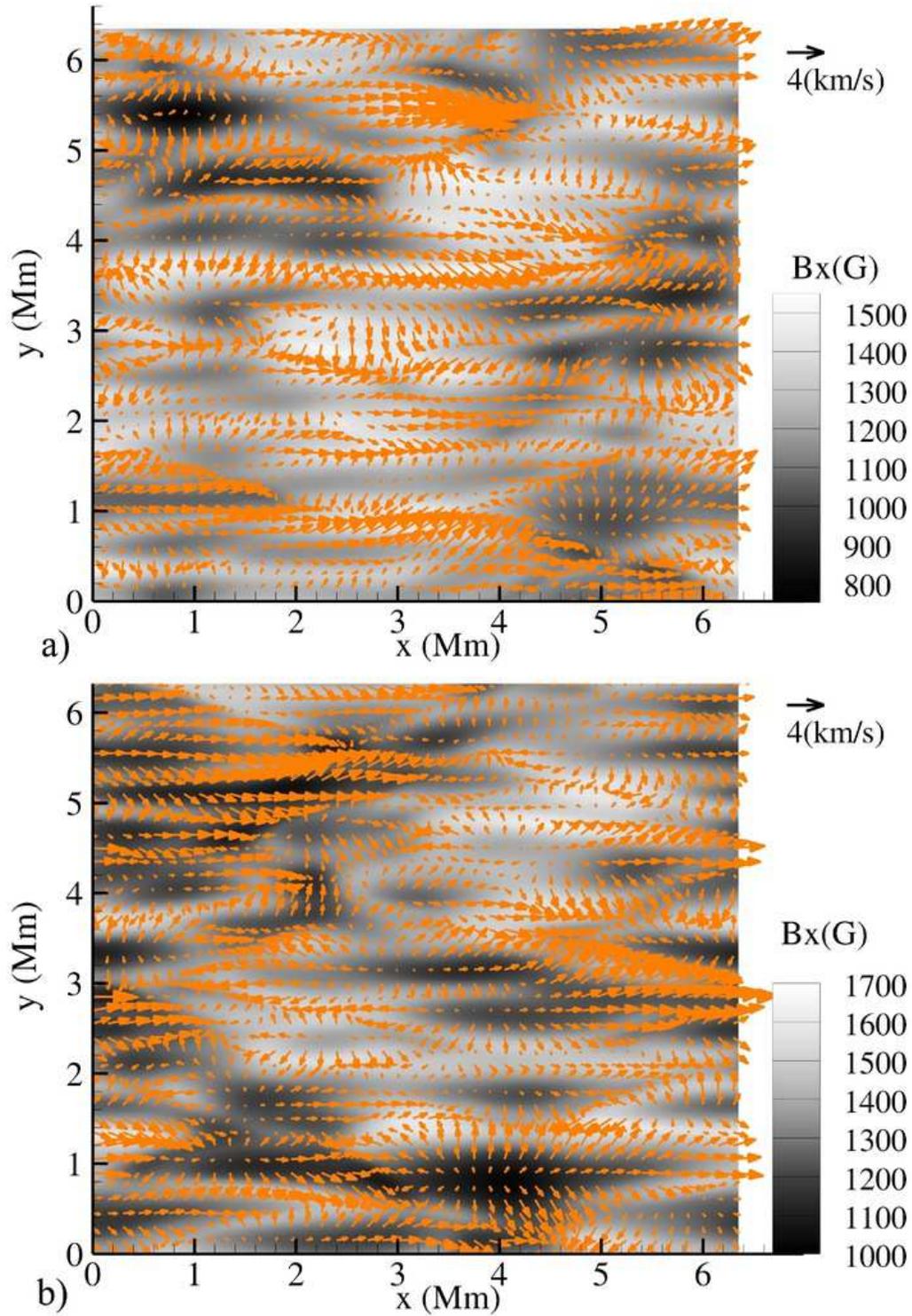}
\caption{The gray-scale maps show snapshots of the magnetic field
component, $B_x$, in the direction of the field inclination, and
arrows show the horizontal velocity at the solar surface for
different initial magnetic field strength, $B_0$: a) 1000
G, b) 1200 G, and the inclination angle, $\alpha=85^\circ$.}
\end{center}
\end{figure}

\begin{figure}
\begin{center}
\includegraphics[width=0.9\linewidth]{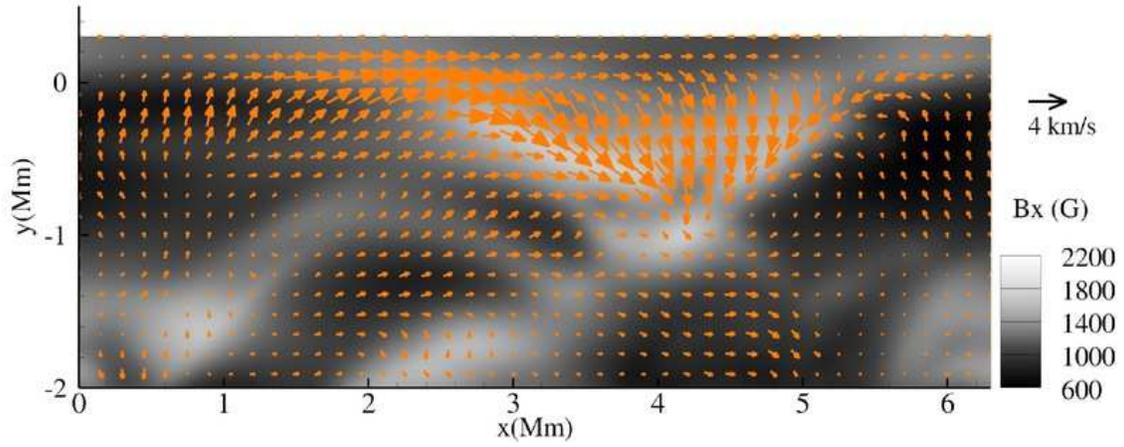}
\caption{The same as Fig.~2 but a vertical $xz$-plane cut for
$B_0=1000$ G and $\alpha=85^\circ$.} \label{spot}
\end{center}
\end{figure}

\begin{figure}
\begin{center}
\includegraphics[width=0.9\linewidth]{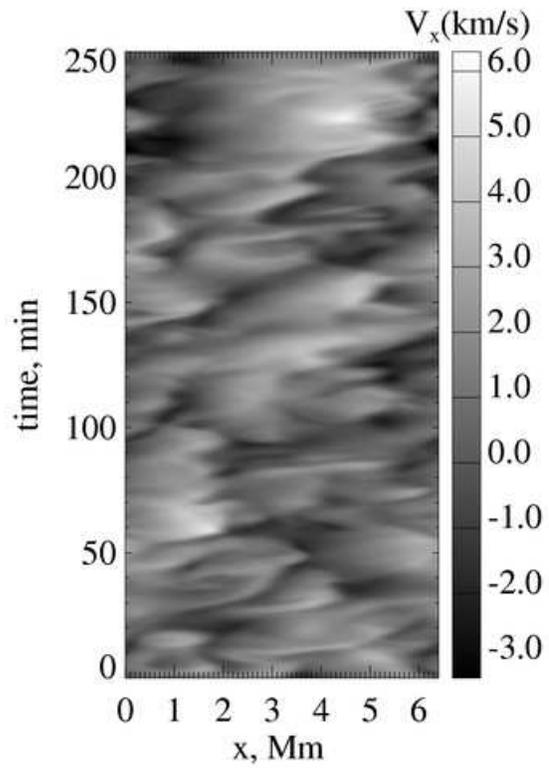}
\caption{Time-space slice of the velocity component, $V_x$, in
the direction of the field inclination for $B_0=1200$ G,
$\alpha=85^\circ$.} \label{spot}
\end{center}
\end{figure}

\begin{figure}
\begin{center}
\includegraphics[width=\linewidth]{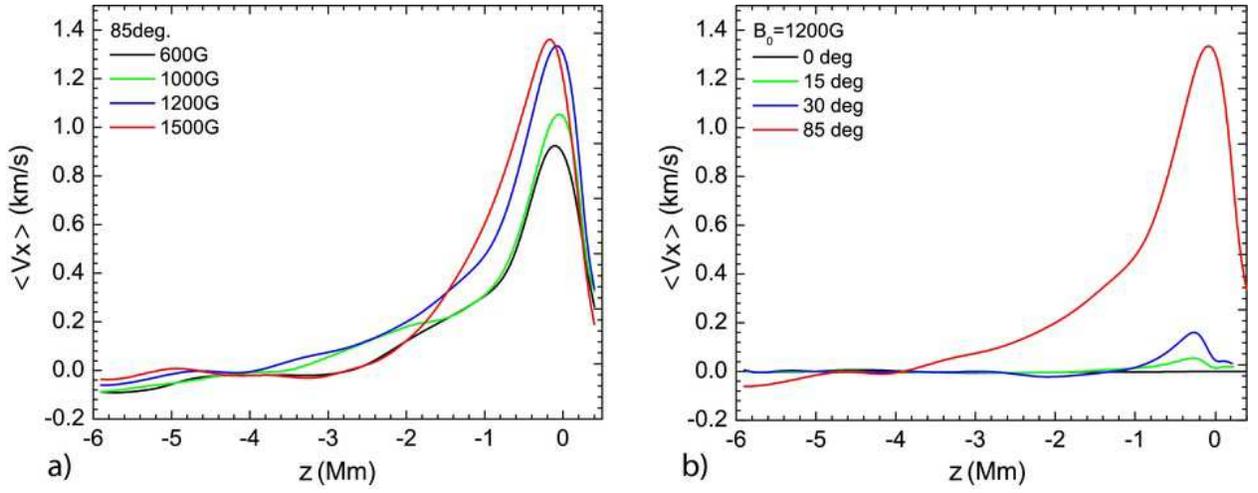}
\caption{Distributions of the averaged horizontal velocity
component, $<V_x>$, with depth, $z$: a) for the magnetic field strength
of 600, 1000, 1200, and 1500 G, and the inclination angle,
$\alpha=85^\circ$; b) $B_0=1200$ G, $\alpha$=0, 15, 30 and 85 degrees.}
\label{spot}
\end{center}
\end{figure}

\end{document}